\documentstyle[12pt]{article}
\renewcommand{\vec}[1]{{\mbox{\boldmath$#1$}}}
\newcommand{\lyxaddress}[1]{
  \par {\raggedright #1
  \vspace{0.3em}
  \noindent\par}
}
\makeatother
\begin{document}
\title{Relativistic calculations of isotope shifts in highly charged
ions}
\author{I.I. Tupitsyn$^{1,2}$, V.M. Shabaev$^{1,2}$,
J.R. Crespo L\'opez-Urrutia$^{2}$, \\
I. Dragani\'c$^{2,3}$, R. Soria Orts$^{2}$,
and J. Ullrich$^{2}$}
\maketitle
\lyxaddress{$^1$
Department of Physics,
 St.~Petersburg State University, Oulianovskaya 1, Petrodvorets,
198504 St.~Petersburg,
Russia}
\lyxaddress{$^2$
Max-Planck Institut f\"ur Kernphysik,
Saupfercheckweg 1, D-69117
Heidelberg, Germany}
\lyxaddress{$^3$
Institute of Nuclear Sciences "Vin\v{c}a", Laboratory of Physics (010),
P.O. Box 522, 11001 Belgrade, Yugoslavia}

\begin{abstract}
The isotope shifts of forbidden transitions in Be- and B-like
argon ions are calculated. It is shown that only using
the relativistic recoil operator  can provide
a proper evaluation of the mass isotope shift, which
strongly dominates over the field isotope shift for the ions
under consideration. Comparing the isotope shifts calculated
with the current experimental uncertainties indicates very good
perspectives for a first test of the relativistic theory of the recoil
effect in middle-Z ions.
\end{abstract}

PACS number(s): 31.30.Jv, 31.30.Gs
%
%
\section{Introduction}

   Recent high-precision measurements of forbidden transitions in
highly charged argon ions \cite{dra03} provide very good
possibilities for tests of relativistic and
quantum electrodynamic (QED) effects
in middle-Z  few-electron systems.
In particular, in the case of Ar$^{13+}$
the QED contribution to the $2s^2\, 2p\;^2P_{1/2}-^2P_{3/2}$
transition energy is by four orders of magnitude larger than the
corresponding experimental error. From the theoretical side,
however, large efforts must be undertaken to achieve this
accuracy (for the current theoretical results, see \cite{dra03}).
The main goal of this paper is to calculate the isotope shifts
of forbiden transitions in Be-like and B-like argon and
to examine a possibility for their experimental determination.

The isotope shifts in the atomic transition energies arise from
the finite nuclear mass (recoil effect) and
from the finite nuclear size (field or volume shift). Generally,
the field shift, caused by the penetration of the electron wave
functions into the nuclear region, dominates for heavy atoms whereas
the mass shift dominates for light atoms.

The relativistic units ($\hbar=c=1$) are used in the paper.
%

\section{Mass shift}

The nonrelativistic (NR) theory of the mass shift for many-electron
atoms was first formulated by Hughes and Eckart \cite{hug30}.
This shift can be represented as the sum of two parts,
the normal mass
shift (NMS) and the specific mass shift (SMS).
The normal mass shift operator is a one-particle operator. It
is given by
\begin{equation}
{H}_{\rm NMS}^{\rm (nr)} \,=\, \frac{1}{2M} \, \sum_{i} \, \vec{p}^2_i,
\label{NMS1}
\end{equation}
where $\vec{p}_i$ is the momentum operator and $M$ is the nuclear mass.
The specific mass shift operator, which is a two-particle operator,
can be written in the form
\begin{equation}
{H}_{\rm SMS}^{\rm (nr)} \,=\, \frac{1}{2M} \, \sum_{i \ne j} \,
 \vec{p}_i \cdot \vec{p}_j.
\label{SMS1}
\end{equation}
The sum of expressions (\ref{NMS1}) and (\ref{SMS1}) defines the
total recoil operator in the nonrelativistic theory.

  The full relativistic theory of the nuclear recoil effect can be
formulated only within quantum electrodynamics. Such a theory was
first formulated in \cite{sha85}, where the complete $\alpha Z$ -
dependence formulas for the recoil corrections to the atomic
energy levels to first order in $m/M$
 were derived. Later, these formulas were rederived
in simpler ways for the case of a hydrogenlike atom
\cite{yel94,pac95} as well as for a general case
of a many-electron atom \cite{sha98a} (see also
\cite{sha02}).
Full relativistic
calculations of the recoil effect, based on these formulas,
were performed for H- and Li-like ions
 \cite{art95,sha98b,sha98c}. In \cite{pac95}, they
were employed to calculate the recoil corrections in
H-like atoms to order $(\alpha Z)^6 m^2/M$,
where $\alpha$ is the fine structure constant.
As follows from these formulas, within the lowest-order relativistic
approximation ($\sim v^2/c^2$) and to first order in $m/M$,
the recoil corrections
can be derived by using the following recoil Hamiltonian
\begin{equation}
{H}_M \,=\, \frac{1}{2M} \, \sum_{i,j} \left[ \vec{p}_i \,
\cdot \vec{p}_j \,-\,
\frac{\alpha Z}{r_i} \left( \vec{\alpha}_i \,+\, \frac{(\vec{\alpha}_i \cdot
\vec{r}_i) \, \vec{r}_i} {r_i^2} \right) \,\cdot\vec{p}_j \right],
\label{HM}
\end{equation}
where  $\vec{\alpha}$ is a vector incorporating the Dirac matrices.
The expectation value of $H_{M}$  on the Dirac wave function
yields the recoil correction  to the atomic energy level
to first order in $m/M$ (here and in what follows,
the Dirac wave functions are
the eigenvectors of the Dirac-Coulomb-Breit Hamiltonian).
An independent derivation of  Hamiltonian (\ref{HM}) was presented in
\cite{pal87}. In \cite{sha94}, this Hamiltonian was employed to calculate
the $(\alpha Z)^4 m^2/M$ corrections to the energy levels of He-
and Li-like middle-Z ions.

According to expression (\ref{HM}), the lowest-order
relativistic correction to the one-electron recoil operator is given by
\begin{equation}
{H}_{\rm NMS}^{\rm (r)}\,
=\, -\frac{1}{2M} \, \sum_{i}
\frac{\alpha Z}{r_i} \left( \vec{\alpha}_i \,+\, \frac{(\vec{\alpha}_i \cdot
\vec{r}_i) \, \vec{r}_i} {r_i^2} \right) \,\cdot\vec{p}_i \,,
\label{RNMS}
\end{equation}
The corresponding two-electron correction is
\begin{equation}
{H}_{\rm SMS}^{\rm (r)} \,
=\,- \frac{1}{2M} \, \sum_{i \ne j}
\frac{\alpha Z}{r_i} \left( \vec{\alpha}_i \,+\, \frac{(\vec{\alpha}_i \cdot
\vec{r}_i) \, \vec{r}_i} {r_i^2} \right) \,\cdot\vec{p}_j \,.
\label{SMS2}
\end{equation}
To the lowest order in $m/M$, the mass isotope shift is determined as
the difference of the expectation values of the recoil Hamiltonian
for two different isotopes:
\begin{eqnarray}
\delta E_{\rm MS} \,=\, \langle \psi |(H_{M_1}-H_{M_2})| \psi \rangle
=-\frac{\delta M}{M_1 \,M_2}K\,,
\label{K1}
\end{eqnarray}
where $\delta M = M_1-M_2$ is the nuclear mass difference,  $|\psi\rangle$
is the eigenvector of the Dirac-Coulomb-Breit Hamiltonian,
and the constant $K$ is defined by
\begin{eqnarray} \label{K2}
K/M=\langle \psi|H_M|\psi\rangle\,.
\end{eqnarray}

It should be stressed here that our approach
to the relativistic recoil effect differs from those
in  \cite{Blundell,Parpia,Safronova,bha01}, where the nonrelativistic recoil
Hamiltonian was evaluated with the Dirac wave functions. Averaging the
nonrelativistic recoil operator with the Dirac wave functions strongly
overestimates the relativistic correction to the recoil effect. It is caused
by the fact (see \cite{sha85,art95}) that the relativistic contribution that
results from averaging
the nonrelativistic nuclear recoil operator  with
the Dirac wave functions is considerably cancelled by the contribution
of the relativistic correction operator defined by equations
(\ref{RNMS}),(\ref{SMS2}).
For this reason,  the relativistic wave functions must be employed only
in calculations with the relativistic recoil Hamiltonian (\ref{HM}).
%

\section{Field shift}
%
The field isotope shift is caused by different nuclear charge
distributions for different isotopes.  For the nuclear charge
distribution we used the Fermi model:
\begin{equation}
\rho(r,R) \,=\, \frac{N}{1+\exp[(r-c)/a]},
\end{equation}
where the parameter $a$ was chosen to be $a=2.3/4 \ln3$ and parameters
$N$ and $c$ can be obtained from the known value of
the root-mean-square (rms) nuclear charge radius
 $R=\langle r^2\rangle^{1/2}$ and from
the normalization condition for $\rho(r,R)$:
\begin{equation}
 \int \,d\vec{r}  \, \rho(r,R) \,=\, 1\,.
\end{equation}
With a high accuracy, $N$ and $c$ can be determined by the
following analytical formulas (see, e.g., \cite{sha93})
\begin{eqnarray}
N&=&\frac{3}{4\pi c^3}\Bigl(1+\frac{\pi^2 a^2}{c^2}\Bigr)^{-1}\,,
\nonumber\\
c&=&\sqrt{\frac{5}{3}\langle r^2\rangle -\frac{7}{3}\pi^2 a^2}\,.
\end{eqnarray}
   The potential of an extended nucleus is given by
\begin{equation}
V_{N}(r,R) \,=\, -4 \pi \alpha Z \,\,
\int \limits_{0}^{\infty} \,
{\rm dr^{\prime}} \,r^2 \, \rho(r^{\prime},R) \,\frac{1}{r_{>}}, \qquad
r_{>} \,=\, {\rm max}(r,r^{\prime}).
\end{equation}
This potential was used in the Dirac-Coulomb-Breit Hamiltonian to obtain
the relativistic wave functions. The field isotope shift can be
determined by the formula
\begin{equation}
\delta E_{\rm FS} \,=\,\langle \psi| \sum_i \, \delta V_{N}(r_i,R)
|\psi\rangle \,,
\end{equation}
where
\begin{equation}
\delta V_N(r) \,=\, V_N(r,R+\delta R) \,-\, V_N(r,R),
\end{equation}
$\delta R$ is the difference of the rms nuclear charge
radii for two isotopes, and $\delta \langle r^2 \rangle$ is the related
 difference of the mean-square nuclear
radii. Within the precision required,
the direct evaluation of the field isotope shift by solving the Dirac-Coulomb-Breit
equation for two different isotopes yields the same results.

Concluding this section, we note that in the case of a hydrogenlike
atom the isotope shift can easily be calculated according to
analytical formulas presented in \cite{sha93}.
%
\section {Method of calculation}
In this work, the large-scale configuration-interaction (CI) Dirac-Fock
(DF) method was used to solve the Dirac-Coulomb-Breit equation
and to calculate the energies and the isotope shifts
of the forbidden transitions in Ar$^{14+}$, Ar$^{13+}$, and Kr$^{22+}$.
The many-electron wave function $\Psi(\gamma J)$ with quantum numbers
$\gamma$ and $J$ was expanded in terms of a large number of the
configuration state functions (CFSs) with the same J
\begin{equation}
\Psi{(\gamma J)} \,=\, \sum_{\alpha} \,c_{\alpha} \, \Phi_{\alpha} (J).
\label{expan1}
\end{equation}
For every relativistic atomic configuration the CFSs $ \Phi_{\alpha} (J)$
are eigenfunctions of the square of total angular momentum $ J^2$ and
they can be obtained as the linear combinations of the Slater
determinants corresponding to this configuration. The set of the CFSs in
expansion (\ref{expan1}) was generated including all single and double
excitations and some part of triple excitations.

The Slater determinants are constructed from one-electron four-component
Dirac spinors (orbitals). For the occupied shells these orbitals
($\varphi_{j}$)
were obtained by the multiconfiguration Dirac-Fock (DF) method. The other
vacant orbitals ($\tilde{\varphi}_{j}$) were obtained by solving the
Dirac-Fock-Sturm equations
\begin{equation}
\left [ {h}^{\rm DF} \,-\,\varepsilon_{j_0}
\right] \,\tilde \varphi_{j}
 \,=\, \lambda_{j} \, W(r) \tilde \varphi_{j},
\label{Sturm1}
\end{equation}
where $ {h}^{\rm DF}$ is the Dirac-Fock operator, $\varepsilon_{j_0}$ is
the one-electron energy of the occupied DF orbital  $\varphi_{j_0}$,
and  $W(r)$ is a constant sign weight function. The parameter $\lambda_{j}$
in  equation (\ref{Sturm1}) can be considered as an eigenvalue of the
Sturmian operator. If  $W(r) \to 0$ at $r \to \infty$, all Sturmian
functions $\varphi_{j}$ have the same asymptotics at $r \to \infty$.
It is clear that
for $\lambda_j=0$ the Sturmian function coincides with
the reference DF orbital $\varphi_{j_0}$.
 The widely known choice of the weight function
is $W(r)=1/r$, which leads to the well known 'charge quantization'. In the
relativistic case this choice of the weight function is not very successful.
In our calculations we used the following weight function
\begin{equation}
W(r)  \,=\, - \,\, \frac{1 \,-\, \exp(-(\alpha \, r)^2)}{(\alpha \, r)^2}\,.
\label{sturm2}
\end{equation}
In contrast to $1/r$, this weight function is regular at origin.
It is well-known that the Sturmian operator is Hermitian and it does
not contain continuum spectra in contrast to the Fock operator.
Therefore, the set of the Sturmian eigenfunctions forms the discrete
and complete basis set of one-electron wave functions.

\section{Results and discussion}

Before discussing our results for the isotope shifts in argon ions,
we examine our calculations of the specific mass shift
 for the $3p_{1/2} - 3s$ transition in Na-like ions
and compare them with the related results from \cite{Safronova}.
In Table 1 we present our results for the SMS
contribution to the isotope-shift
constant $K$, defined by equation (\ref{K2}).
The second column shows the nonrelativistic values of
this contribution.
These values were obtained  within the
 same CIDF method and the same computer
code, as described above,
by the 1000-times increase of the velocity of light
(in atomic units).
We verified this nonrelativistic
limit by comparing the total energies for the Dirac-Coulomb-Breit
Hamiltonian with the total energies obtained by the fully
nonrelativistic method based on the Schr\"odinger
Hamiltonian and
on the same calculation scheme. The total energies
exactly coincide with each other for all the ions.
In the third column we present the results obtained
by averaging the nonrelativistic recoil operator  (\ref{SMS1}) on the
Dirac wave functions. The fourth column indicates the relativistic correction
that results from  using the Dirac wave functions in the calculation
of the nonrelativistic-recoil-operator contribution. In the fifth column
we give the
relativistic correction obtained by averaging the relativistic
correction operator
 (\ref{SMS2}). In the sixth column the total relativistic data are listed.
The last column indicates the results of \cite{Safronova}, based on averaging
the nonrelativistic recoil operator with the Dirac wave functions.
Comparing the relativistic corrections, presented in the fourth and fifth
columns, we observe that the absolute value of the second relativistic correction
is larger than the absolute value of the first one and, as a result,
the total relativistic correction and the correction that accounts only for the
relativistic effects on the wave functions are of opposite sign.
This means that using only
the nonrelativistic part of the recoil operator in calculations based on employing
the Dirac wave functions gives strongly incorrect results
for the relativistic recoil effect.

For the $3p_{1/2} - 3s$ transition in
neutral Na, the experimental value of the isotope shift
$^{22}$Na/$^{23}$Na
amounts to  -758.5(0.7) MHz \cite{Pescht}.
To derive the corresponding SMS
we have to subtract the NMS and FS
from the experimental isotope shift.
The nonrelativistic NMS has been evaluated as the
$-m(1/M_1-1/M_2)$ fraction of
the nonrelativistic transition energy, obtained as the difference
between the experimental value  of the transition energy
 (16956.2~cm$^{-1}$) and the related
relativistic correction (45.4~cm$^{-1}$),
calculated in this work. As a result, the nonrelativistic NMS
was obtained to be
  -547.5~MHz.
The sum of this value and the relativistic NMS correction
(-~9.7~MHz), calculated in this work, yields
   -557.2 MHz
for the total NMS value. Subtracting this NMS value and our
theoretical FS value (6.9 MHz) from the experimental isotope shift,
we obtain the value
-208.2~MHz
for the SMS and
-105.8~(GHz amu)
for the SMS  contribution to the constant $K$. This value differs
from our theoretical result
 -98.5~(GHz amu) (Table 1) by 7\%
 and from the result of \cite{Safronova} by 9\%.
The disagreement between the theoretical and experimental
values is mainly caused by the correlation effects,
which are extremely large for neutral Na.
It is very difficult to compute the isotope shift
parameters for neutral Na with  high accuracy.
For systematic investigations of the nonrelativistic isotope
shift by  the  multiconfiguration
Hartree-Fock method we refer to \cite{Fischer},
where an extremly large CI expansion was used.
Although in our relativistic calculations the CI
expansion is not so large, it  is sufficient to
 show the role of the relativistic effects in
the series of Na-like ions. Adding our relativistic correction
to the nonrelativistic contribution from \cite{Fischer} 
yields $K=-103.8$ (GHz amu) that is in a fair agreement with
the experimental value presented above.

  Table 2 shows the field-shift constant $F$, defined by
$\delta E_{\rm FS}=-F\delta\langle r^2\rangle $,
for the $3p_{1/2} - 3s$
transition in Na-like ions. The comparison of the nonrelativistic
(second column)
and  relativistic (third column) data demonstrates the role of
the relativistic effects
in the field isotope shifts. Our relativistic field-shift constants are
in a good agreement with the related data from \cite{Safronova},
presented in the fourth
column of the table.

In Table 3 we present the results of our calculations for
the isotope shifts in  $^{40}$Ar$^{13+}$/$^{36}$Ar$^{13+}$,
 $^{40}$Ar$^{14+}$/$^{36}$Ar$^{14+}$, and
$^{86}$Kr$^{22+}$/$^{84}$Kr$^{22+}$.  To calculate the field
isotope shift, we used the rms nuclear charge
radii given in \cite{fri95}. Since we consider transitions
between levels which differ from each other only by
the total angular momentum, the corresponding transition energies
are completely determined by relativistic and QED effects.
In particular, it means that all the mass and field shift contributions
given in the table are of pure relativistic origin.
Comparing the individual mass shift contributions given in the Table 3,
we again observe a very significant cancellation of the
contributions from the nonrelativistic part of the recoil operator
($H_{\rm NMS}^{\rm (nr)}+H_{\rm SMS}^{\rm (nr)}$)
by the corresponding contributions from the relativistic part of the recoil
operator ($H_{\rm NMS}^{\rm (r)}+H_{\rm SMS}^{\rm (r)}$).
It can also be seen that the mass isotope shift strongly
dominates over the field isotope shift for the ions under
consideration.

In Table 4, we compare the isotope shifts calculated in this paper
with the total theoretical and experimental
values for the transition energies under consideration
\cite{dra03}.
The non-QED parts of the theoretical values for the  transition energies
were obtained by the large-scale CIDF method described in the
previous section. The QED corrections were evaluated by using
the one-electron Lamb shift data taken from \cite{joh88} with an effective
nuclear charge number $Z_{\rm eff}$. For a given one-electron state,
$Z_{\rm eff}$ was chosen to reproduce the related DF electron charged
density at the Compton wavelength distance from the nucleus.
For the argon ions, these results are also in a fair
agreement with the corresponding results
of Refs. \cite{saf96a,saf96b} presented in the third column.
  In the sixth column of Table 4 we present the total isotope shifts in
$^{40}$Ar$^{13+}$/$^{36}$Ar$^{13+}$. These data were obtained
with the Dirac multiconfiguration wave functions
 and their relative errors are determined
by the relative errors in the wave functions.
As one can see from the table, the isotope shift
in $^{40}$Ar$^{13+}$/$^{36}$Ar$^{13+}$ is by an order of
magnitude larger than the current experimental error.
This provides very good perspectives for a first test
of the relativistic theory for the recoil effect in
middle-Z ions.

\section*{Acknowledgments}

Valuable conversations with V.P. Shevelko and H. Tawara are
gratefully acknowledged.
We acknowledge support from Max-Planck-Institut
f\"ur Kernphysik in Heidelberg, from RFBR (Grants Nos
01-02-17248 and 00-03-33041), from the program "Russian Universities"
(Grant No. UR.01.01.072), and from the Russian Ministry of
Education (Grant No. E02-3.1-49).

\clearpage
{
\begin{table}
%
\caption{The specific-mass-shift contribution to the
isotope shift constant $K$ (GHz amu) for
the $3p_{1/2} - 3s$ transition in Na-like ions.}
%
\begin{center}
\begin{tabular}{|l|c|c|c|c|c|c|}
\hline
 & & & & & & \\[-12pt]
 Ion & NSMS &  SMS & SMS - NSMS & RSMS & Total  &  SMS \cite{Safronova} \\ [2mm]
\hline
  {Na}         &    -100.1   &   -101.0 &-0.9 &    2.5  &    -98.5  &    -97  \\
 {Mg$^{1+}$}   &    -412.7   &   -416.1  & -3.4 &  10.0  &   -406.1  &   -362  \\
 {Al$^{2+}$}   &    -926.3   &   -934.8 &-8.5 &   24.0  &   -910.8  &   -837  \\
 {Cl$^{6+}$}   &   -5082.9   &  -5162.4 &-79.5 &  183.2  &  -4979.3  &  -4846  \\
 {Ar$^{7+}$}   &   -6648.2   &  -6765.7  &-117.5&  261.6  &  -6504.1  &  -6363  \\
 {Fe$^{15+}$}  &   -26677   &    -27713  &-1036& 1995.3  & -25718.6  & -25662  \\
 {Xe$^{43+}$}  &  -200802    &  -240996  &-40194 & 69295  &  -171701  &     -   \\
\hline
\multicolumn{6}{c}{}\\[-5pt]
\multicolumn{6}{l}{NSMS - nonrelativistic specific-mass shift} \\
\multicolumn{6}{l}{SMS - specific-mass shift calculated with the Dirac
 wave function} \\
\multicolumn{6}{l}{RSMS - relativistic-operator correction to the specific mass shift } \\
\multicolumn{6}{l}{Total - SMS+RSMS } \\
\end{tabular}
\end{center}
\end{table}
%
\clearpage
{
\begin{table}
%
\caption{Field shift constant $F$ (MHz/fm$^2$) for
the $3p_{1/2} - 3s$ transition in Na-like ions.
$\langle r^2 \rangle ^{1/2}~ =~ 0.836 \, A^{1/3} + 0.570$ ($A$ is the atomic mass number). }
%
\begin{center}
\begin{tabular}{|l|c|c|c|}
\hline
&&&\\[-12pt]
 Ion & NFS &  FS  & FS \cite{Safronova}
\\ [2mm]
\hline
&&&\\
  {Na}         &      34.6   &     36.45   &  38.4  \\
 {Mg$^{1+}$}   &    115.9   &     123.2    &  125.8  \\
 {Al$^{2+}$}   &    247.3   &     265.3    &  268.4  \\
 {Cl$^{6+}$}   &   1538.8   &    1726.2    &  1729.7  \\
 {Ar$^{7+}$}   &   2133.8   &   2423.4    &  2426.5  \\
 {Fe$^{15+}$}  &  14095.9   &  18107.9    &  18115  \\
 {Xe$^{43+}$}  &  373760   & 965738    &     -  \\
\hline
\multicolumn{4}{c}{}\\[-5pt]
\multicolumn{4}{l}{NFS - nonrelativistic field shift constant} \\
\multicolumn{4}{l}{FS - relativistic field shift constant} \\
\end{tabular}
\end{center}
\end{table}
%
%
\clearpage
{
\begin{table}[t]
%
%
\caption{Individual contributions to the isotope shifts
of the forbidden transition in
$^{40}$Ar/$^{36}$Ar and $^{86}$Kr/$^{84}$Kr (cm$^{-1}$).
The rms nuclear charge radii used in the calculation are
$\langle r^2 \rangle^{1/2}$ = 3.390, 3.427, 4.188, and 4.184 fm
for $^{36}$Ar, $^{40}$Ar, $^{84}$Kr, and $^{86}$Kr, respectively
\cite{fri95}.}
%
\begin{center}
\begin{tabular}{|c|c|c|c|c|c|c|c|}
\hline
&&&&&&&\\
 Ion &  Transition   & NMS  &  SMS   & RNMS & RSMS & FS & Total \\
\hline
&&&&&&&\\
 { Ar$^{ 13+}$}  & { 2s$^{ 2}$2p$^{ 1}$ $^{ 2}
{\rm P_{ 1/2} - ^2P_{3/2}}$ }
  & 0.1053  & -0.0742 & -0.0822 & 0.1151 &  -0.0005 & 0.0635 \\
 { Ar$^{ 14+}$}  & { 2s$^{ 1}$2p$^{ 1}$
$^{ 3}{\rm  P_{ 1} - ^3P_{2}}$ }
  & 0.0797  & -0.0698 & -0.0627 & 0.0887 &  -0.0001 & 0.0358 \\
 {Kr$^{ 22+}$}  & { 3s$^{ 2}$3p$^{ 2}$
$^{ 3}
{\rm P_{1} - ^3P_{2}}$ }
  & 0.0053  & 0.0010 & -0.0025 & 0.0000 &   0.0001 & 0.0039 \\
\hline
\multicolumn{8}{c}{}\\[-5pt]
\multicolumn{8}{l}{NMS - normal mass shift calculated with the Dirac wave function } \\
\multicolumn{8}{l}{SMS - specific mass shift calculated with the Dirac wave function } \\
\multicolumn{8}{l}{RNMS - relativistic-operator correction to the normal mass shift } \\
\multicolumn{8}{l}{RSMS - relativistic-operator correction to the specific mass shift } \\
\multicolumn{8}{l}{FS - field shift } \\
\end{tabular}
\end{center}
\end{table}
%
\clearpage
{
\begin{table}
\caption{Energies  of forbidden transitions for Ar and Kr ions
 and the isotope shifts in $^{40}$Ar/$^{36}$Ar and $^{86}$Kr/$^{84}$Kr.}
%
\small
\begin{center}
\begin{tabular}{|c|c|c|c|c|c|}
\hline
&&&&\\[-12pt]
 Ion &  Transition   & Theory \cite{saf96a,saf96b}
& Theory \cite{dra03}, this work
  &  Experiment \cite{dra03} & Isotope
\\ 
     &              & $\lambda$ (nm, air)  & $\lambda$ (nm, air)
 & $\lambda$ (nm, air)  & shifts (nm) \\ [2mm]
\hline
&&&&&\\
 { Ar$^{13+}$}  & { 2s$^{ 2}$2p$^{1}$ $^{ 2}
{\rm P_{1/2} \,-\, ^2P_{3/2}}$ }
  & 440.99 & 441.16(27)  & 441.2559(1) & 0.00126 \\
 {Ar$^{ 14+}$}  & {2s$^{ 1}$2p$^{ 1}$
 $^{ 3}
{\rm  P_{1} \,-\, ^3P_{2}}$ }
  & 593.88  & 594.24(30)  & 594.3880(3) & 0.00136 \\
 { Kr$^{ 22+}$}  & { 3s$^{ 2}$3p$^{ 2}$
 $^{ 3}
{\rm  P_{ 1} \,-\, ^3P_{2}}$ }
  &  - & 383.35(95)  & 384.1146(2) & 0.00005 \\
\hline
\end{tabular}
\end{center}
\normalsize
\end{table}
%
%
\clearpage
\newpage

%

\begin{thebibliography}{99}
%
\bibitem{dra03}
I. Dragani\'c, J.R. Crespo L\'opez-Urrutia, R. DuBois, S. Fritzsche,
               V.M. Shabaev, R. Soria Orts, I.I. Tupitsyn. Y. Zou,
 and J. Ullrich, to be published.
%
\bibitem{hug30} D.S. Hughes and C. Eckart, Phys. Rev. {\bf 36},
                 694 (1930).
%
\bibitem{sha85} V.M. Shabaev, Teor. Mat. Fiz. {\bf 63}, 394 (1985)
[Theor. Math. Phys. {\bf 63}, 588 (1985)];
   Yad. Fiz. {\bf 47}, 107 (1988)
                 [Sov. J. Nucl.Phys. {\bf 47}, 69 (1988)].
\bibitem{yel94}
A.S. Yelkhovsky, Preprint BINP 94-27
(Budker Inst. of Nuclear Physics, Novosibirsk, 1994);
hep-th/9403095 (1994).
\bibitem{pac95}
K. Pachucki and H. Grotch, Phys. Rev. A {\bf 51}, 1854 (1995).
\bibitem{sha98a}
V.M. Shabaev, Phys. Rev. A {\bf 57}, 59 (1998).
\bibitem{sha02}
V.M. Shabaev, Phys. Rep. {\bf 356}, 119 (2002);
In {\it The Hydrogen Atom}, edited by
S.G. Karshenboim et al., (Springer, Berlin, 2001), p. 714.
\bibitem{art95}
A.N. Artemyev, V.M. Shabaev, and V.A. Yerokhin,
Phys. Rev. A {\bf 52}, 1884 (1995);
 J. Phys. B {\bf 28}, 5201 (1995).
\bibitem{sha98b}
V.M. Shabaev, A.N. Artemyev, T. Beier, G. Plunien,
V.A. Yerokhin, and  G. Soff, Phys. Rev. A  {\bf 57}, 4235
(1998);  Phys. Scr. T {\bf 80}, 493 (1999).
\bibitem{sha98c}
V.M. Shabaev, A.N. Artemyev, T. Beier, and G. Soff,
J. Phys. B {\bf 31}, L337 (1998).
%
\bibitem{pal87} C.W. Palmer, J. Phys.B {\bf 20},
                   5987 (1987).
\bibitem{sha94}
V.M. Shabaev and A.N. Artemyev, J. Phys. B {\bf 27}, 1307 (1994).
%
\bibitem{Blundell} S.A. Blundell, Phys. Rev. A
                  {\bf 46}, 3762 (1992).
%
\bibitem{Parpia} F.A. Parpia, M. Tong, and C.F. Fischer, Phys. Rev. A
                  {\bf 46}, 3717 (1992).
%
\bibitem{Safronova} M.S. Safronova and W.R. Johnson, Phys. Rev. A
                  {\bf 64}, 052501 (2001).
\bibitem{bha01}
M.I. Bhatti, M. Bucardo, and W.F. Perger, J. Phys. B {\bf 34},
223 (2001).
%
\bibitem{sha93} V.M. Shabaev, J. Phys. B {\bf 26}, 1103 (1993).
%
\bibitem{Pescht} K. Pescht, H. Gerhardt, and E. Matthias, Z. Phys. A
                 {\bf 281}, 199 (1977).
%
\bibitem{Fischer} C. Frose Fischer, P. J\"onsen, M. Godefroid,
                  Phys. Rev. A {\bf 57}, 1753 (1998).
%
\bibitem{fri95}
G. Fricke, C. Bernhardt, K. Heilig, L.A. Schaller, L. Schellenberg,
E.B. Shera, C.W. de Jager, At. Data Nucl. Data Tables {\bf 60},
177 (1995).
\bibitem{joh88}
W.R. Johnson and G. Soff, At. Data and Nucl. Data Tables {\bf 33},
405 (1985).
\bibitem{saf96a} M.S. Safronova, W.R. Johnson, and U.I. Safronova,
Phys. Rev. A {\bf 54}, 2850 (1996).
\bibitem{saf96b} M.S. Safronova, W.R. Johnson, and U.I. Safronova,
Phys. Rev. A {\bf 53}, 4036 (1996).
%
\end{thebibliography}
\end{document}